\documentclass[aps,showpacs,twocolumn]{revtex4}
\usepackage{mathrsfs}
\usepackage{amssymb}
\usepackage{amsmath}
\usepackage{bm}
\usepackage{graphics}
\usepackage{natbib}

\begin{document}
\title{Influence of $\mathcal{PT}$-symmetric complex potentials on the decoupling mechanism in quantum transport process}
\author{Lian-Lian Zhang}
\author{Wei-Jiang Gong}\email{gwj@mail.neu.edu.cn}
\author{Guang-Yu Yi}
\author{An Du}

\affiliation{College of Sciences, Northeastern University, Shenyang
110819, China}
\date{\today}

\begin{abstract}
We consider one system in which the terminal dots of a one-dimensional quantum-dot chain couple equally to the left and right leads and study the influence of $\mathcal{PT}$-symmetric complex potentials on the quantum transport process. It is found that in the case of the Hermitian Hamiltonian, remarkable decoupling and antiresonance phenomena have an opportunity to co-occur in the transport process. For the chains with odd(even) dots, all their even(odd)-numbered molecular states decouple from the leads. Meanwhile, antiresonance occurs at the positions of the even(odd)-numbered eigenenergies of the sub-chains without terminal dots. When the $\mathcal{PT}$-symmetric complex potentials are introduced to the terminal dots, the decoupling phenomenon is found to transform into the Fano antiresonance. In addition, it shows that appropriate magnetic flux can interchange the roles of the odd and even molecular states. These results can assist to understand the quantum transport modified by the $\mathcal{PT}$ symmetry in non-Hermitian discrete systems.

\end{abstract}
\pacs{73.23.Hk, 73.50.Lw, 85.80.Fi} \maketitle

\bigskip

\section{Introduction}
During the past decades, non-Hermitian Hamiltonians have attracted extensive interests. It has been found to be able to exhibit
entirely real spectra if these Hamiltonians have
parity-time ($\mathcal{PT}$) symmetry\cite{Bender1}. Based on this reason, numerous $\mathcal{PT}$-symmetric
systems have been explored in various fields, including the complex
extension of quantum mechanics\cite{Bender3,Mosta}, the quantum
field theories and mathematical physics\cite{Bender4}, open quantum
systems\cite{Rotter}, the Anderson models for disorder systems\cite{Gold,Hei,Moli},
the optical systems with complex refractive indices\cite{Klaiman,Xu,Kottos,Makris,Luo}, and the topological insulators\cite{Hu,Zhu}. In addition, encouraged by the experimental achievement in optical waveguides\cite{Guo,Kip}, optical lattices\cite{Miri}, and in a pair of coupled
LRC circuits\cite{Li}, the non-Hermitian lattice models with $\mathcal{PT}$ symmetry have also received much attention. Moreover, it has been reported that these models can be realized in the Gegenbauer-polynomial quantum chain\cite{Zo}, one-dimensional $\mathcal{PT}$-symmetric chain with disorder\cite{Bendix}, the chain model with two conjugated imaginary potentials at two end
sites\cite{Jin}, the tight-binding model with position-dependent
hopping amplitude\cite{Jogle}, and the time-periodic $\mathcal{PT}$-symmetric
lattice model\cite{Valle}.
\par
For the physics properties of the systems with $\cal PT$-symmetric non-Hermitian Hamiltonians, researchers are used to focusing on their $\cal PT$ phase diagrams as well as the signatures of $\cal PT$-symmetry breaking. It should be noticed that the transport behaviors are also important aspects for the description of physics properties of low-dimensional systems\cite{Datta}. During the past years, lots of interesting transport phenomena have been observed in the quasi-one-dimensional systems, such as the Fano resonance\cite{Fano}, decoupling state\cite{Decouple}, and Kondo resonance\cite{Kondo}. With respect to the Fano resonance, it is characterized by a sharp asymmetric profile in the transmission profile, caused by the quantum interference between the discrete energy level and
the continuum spectrum\cite{Fano2}. For the decoupling phenomenon, it is manifested as the peak disappearance in the transmission spectrum. Thus, when investigating the systems with $\cal PT$-symmetric non-Hermitian Hamiltonians, their transport properties are worthy to make further studies. Recently, researchers began to be concerned in the influence of $\mathcal{PT}$-symmetric complex potentials on these typical quantum transport phenomena. It has been reported that for a Fano-Anderson system, the $\mathcal{PT}$-symmetric imaginary potentials can lead to some pronounced effects on transport properties of Fano systems, including changes from the perfect reflection to perfect transmission, and rich behaviors for the absence or existence of the perfect reflection at one and two resonant frequencies\cite{PRA1}. Following these results, other groups investigated a non-Hermitian Aharonov-Bohm ring system with a quantum dot(QD) embedded in each of its two arms, by considering the complex QD levels. As a result, they found that with appropriate parameters, the asymmetric Fano profile will show up in the conductance spectrum just by non-Hermitian quantity in this system\cite{Lvrong}. These works indeed show the interesting transport properties of the non-Hermitian systems. However, it should be noticed that due to the abundant quantum transport phenomena in low-dimensional systems, more researches should be performed in order to clearly describe the transport properties of the non-Hermitian systems.
\par
In the present work, we would like to study the influence of $\mathcal{PT}$-symmetric complex potentials on the decoupling mechanism in the quantum transport process. For this purpose, we consider one system in which the terminal QDs of a one-dimensional QD chain couple to the left and right leads. It is found that by adjusting the structural parameters, remarkable decoupling phenomenon has an opportunity to take place, accompanied by the occurrence of antiresonance. We see that for the chains of odd(even) QD, all their even(odd)-numbered molecular states decouple from the leads, and antiresonance occurs at the positions of the even(odd)-numbered eigenenergies of the sub-chains without terminal QDs. When the $\mathcal{PT}$-symmetric complex potentials are introduced to the terminal QDs, the decoupling phenomenon will transform into the Fano antiresonance, whereas the previous antiresonance survives. This result exactly means that the $\mathcal{PT}$-symmetric complex potentials play a special role in modifying the decoupling mechanism in the quantum transport process.
\begin{figure}
\centering \scalebox{0.45}{\includegraphics{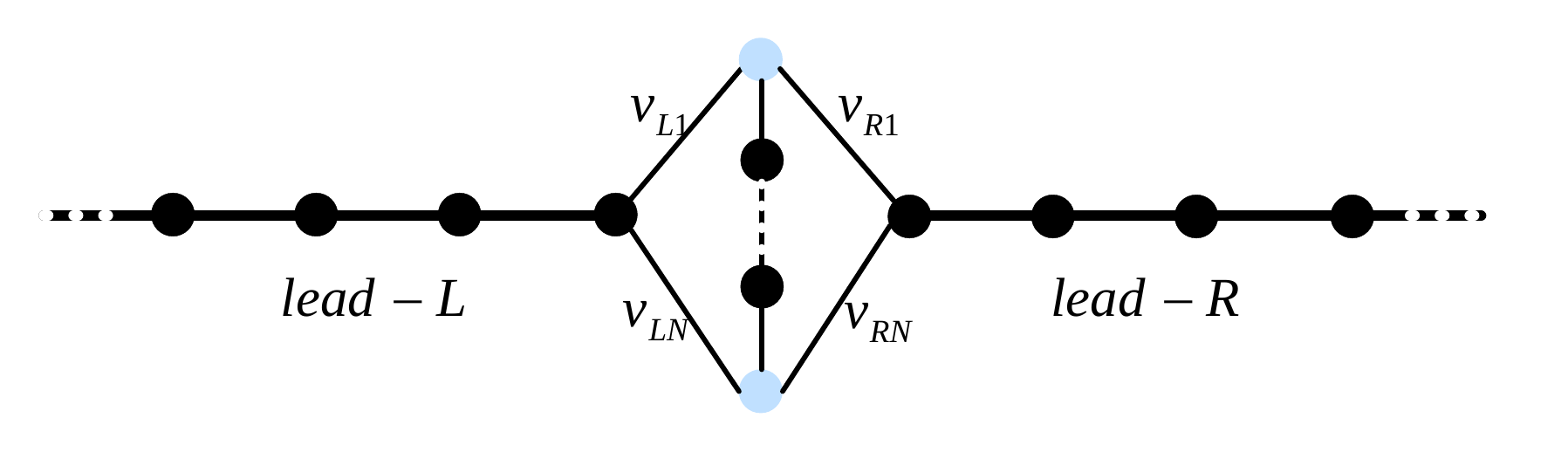}} \caption{
Schematic diagram of one non-Hermitian parallel mesoscopic circuit, where the terminal QDs of a QD chain couple to two leads. The terminal QDs are influenced by $\cal PT$-symmetric complex chemical potentials. The couplings between the terminal QDs and the leads are defined by $v_{\alpha j}$ ($\alpha\in L,R$ and $j=1,N$). \label{structure}}
\end{figure}
\section{Theoretical model}
The structure that we consider is shown in Fig.\ref{structure}, in which the terminal QDs of a QD chain couple to two metallic leads, respectively. We add imaginary potentials to the two terminal QDs of the chain to represent the physical gain or loss during
the interacting processes between the
environment and it. The Hamiltonian of this system can be written as
\begin{eqnarray}
H=\sum_\alpha H_\alpha +H_c+\sum_\alpha H_{\alpha T},
\end{eqnarray}
with its each part given by
\begin{eqnarray}
H_\alpha&=&\sum_{j=1}^{\infty}t_{0}c^{\dag}_{\alpha j}c_{\alpha,j+1}+h.c. \notag\\
H_c&=&\sum_{l=1}^N E_ld_{l}^{\dag}d_{l}+\sum_{l=1}^{N-1}t_ld_{l+1}^{\dag}d_{l}+h.c.,\notag\\
H_{\alpha T}&=&v_{\alpha1}c_{\alpha1}^{\dag}d_{1}+v_{\alpha N}c_{\alpha1}^{\dag}d_{N}+h.c..\label{Hamilton}
\end{eqnarray}
$d^\dag_l$ ($d_l$) is the creation (annihilation)
operator for QD-$l$ of the QD chain with energy level $E_l$. When $E_l$ are real, the Hamiltonian
is Hermitian, whereas if one of them is complex,
the Hamiltonian will become non-Hermitian. $c^\dag_{\alpha j}$
($c_{\alpha j}$) is to create (annihilate) a fermion at the $j$-th site of lead-$\alpha$ with $t_0$
being the hopping amplitude between the nearest sites. $v_{\alpha 1(N)}$ is the tunneling amplitude between
the first ($N$-th) QD in the QD chain and lead-$\alpha$. It is known that in discrete systems, $\cal P$ and $\cal T$ are defined as the space-reflection (parity) operator and the time-reversal operator. If a Hamiltonian obeys the commutation relation $[{\cal PT}, H]=0$, it will be said to be $\cal PT$ symmetric. In our considered geometry, the effect of the $\cal P$ operator is to let ${\cal P}d_{N+1-l}{\cal P}=d_l$ with the linear chain as the mirror axis, and the effect of
the $\cal T$ operator is ${\cal T} i{\cal T}=-i$. Thus, it is not difficult to find that the
Hamiltonian is invariant under the combined operation ${\cal PT}$, under the condition of $t_l=t_c$, $v_{\alpha 1}=v^*_{\alpha' N}$, and $E_l=E^*_{N+1-l}$.

\par
In order to study the quantum transport through this
structure, the transmission function in this system should be
calculated. According to the previous works, various methods can be employed to calculate the transmission function. In this work, we would like to choose the nonequilibrium Green function technique to perform calculations. With the help of such a technique, the transmission function is expressed as\cite{GF1,GF2}
\begin{equation}
T(\omega)=\mathrm{Tr}[\Gamma^L G^a(\omega)\Gamma^R
G^r(\omega)].\label{conductance}
\end{equation}
$\Gamma^{\alpha}=i(\Sigma_{\alpha}-\Sigma^\dag_{\alpha})$ denotes the
coupling between lead-$\alpha$ and the device region. $\Sigma_{\alpha}$, defined as $\Sigma_{jl,\alpha}=v^*_{\alpha j}g_\alpha v_{\alpha l}$, is the selfenergy caused by the coupling between the
QD chain and lead-$\alpha$. $g_\alpha$ is the Green function of the end site of the semi-infinite lead-$\alpha$. Due to the uniform intersite coupling in lead-$\alpha$, the analytical form of $g_\alpha$ can be written out, i.e., $g_\alpha=g_0={\omega \over 2t_0^2}-{\rho_0\over2}i$ with $\rho_0=\sqrt{4t_0^2-\omega^2}/t_0^2$\cite{Self}. In addition, in Eq.(\ref{conductance}) the
retarded and advanced Green functions in Fourier space are involved.
They are defined as $G^r=[\omega+i0^+-H_c-\Sigma_L-\Sigma_R]^{-1}$ and $G^a=[G^r]^{-1}$. By a
straightforward derivation, we can obtain the matrix form of the retarded
Green function, i.e.,
\begin{eqnarray}
G^r(\omega)=\left[\begin{array}{ccccc} g_{1}^{-1}&-t_1^*&0&\cdots & -\Sigma_{1N}\\
-t_1 &g_{2}^{-1}& -t_2^*&0&\vdots\\
& &\ddots& &\\
\vdots &0&-t^*_{N-2}& g_{N-1}^{-1}&-t^*_{N-1}\\
-\Sigma_{N1}&\cdots&0&-t_{N-1}& g_{N}^{-1}\\
\end{array}\right]^{-1}\ \label{green}.
\end{eqnarray}
$g_{l}=\omega+i0^+-E_l-\Sigma_{ll}(\delta_{1l}+\delta_{Nl})$ is the zero-order Green function of QD-$l$
of the QD chain $\Sigma_{jl}=\Sigma_{jl}^L+\Sigma_{jl}^R$.
\par
In our structure, the terminal QDs of the QD chain couple to the two metallic leads simultaneously, and then a quantum ring comes into being [See Fig.1]. Therefore, $v_{\alpha 1(N)}$ has an opportunity to become complex when a local magnetic flux is introduced to thread through such a system. Under gauge transformation, these phase factors can be allocated differently. It is known that for one Hermitian Hamiltonian, the physical variables will not be influenced by these differences. We would like to point out that this physical picture also
applies to our non-Hermitian system. In this work, we choose the symmetric gauge with the phase factor distributed averagely to the four tunneling amplitudes
$v_{L1}=|v_{L1}|e^{i\phi/4}$, $v_{LN}=|v_{LN}|e^{-i\phi/4}$,
$v_{R1}=|v_{R1}|e^{-i\phi/4}$, and $v_{RN}=|v_{RN}|e^{i\phi/4}$.

\section{Analysis about decoupling mechanism}

\par
The transmission function profile reflects the eigenenergy spectrum of the
``molecules" made up of QDs. In other words, each resonant peak in the transmission function spectra represents an eigenenergy of the total molecules, rather than the levels of the individual
QDs. Therefore, it is necessary to transform the Hamiltonian into the molecular orbital representation of the QD chain. It is quite helpful to analyze the numerical results for the transmission function spectra.
\par
We then introduce the electron creation(annihilation) operators
corresponding to the molecular orbits, i.e., $f_{m}^\dag\;
(f_{m})$. By diagonalizing the single-particle Hamiltonian of the QD chain, we find the relation between the molecular and atomic representations (here each site is regarded as an ``atom"),
which is expressed as $[\bm{f}^\dag]=[\bm{\eta}][\bm{d}^\dag]$. The $N\times
N$ transfer matrix $[\bm{\eta}]$ consists of the eigenvectors of the
QD-chain Hamiltonian. In the molecular orbital representation, the
single-particle Hamiltonian takes the form:
$H=\underset{k\alpha\in L,R}{\sum }\varepsilon _{\alpha
k}c_{\alpha k}^\dag c_{\alpha k}+\sum_{m=1}e
_{m}f_{m}^\dag f_{m}+\underset{\alpha k}{\sum }
w_{\alpha m}f_{m}^\dag c_{\alpha k}+{\mathrm {h.c.}}$,
in which $e_m$ is the eigenenergy of the QD chain; $w_{\alpha
m}=v_{\alpha 1}[\bm\eta]^\dag_{1m}+v_{\alpha
N}[\bm\eta]^\dag_{Nm}$, denotes the coupling between
molecular state $|m\rangle$ of the QD chain and state $|k\rangle$ in
lead-$\alpha$. In the case of uniform chain-lead coupling where $|v_{\alpha 1(N)}|\equiv v_0$, the above
relation can be rewritten as
\begin{equation}
w_{\alpha
m}=v_0([\bm\eta]^\dag_{1m}e^{i\phi/4}+[\bm\eta]^\dag_{Nm}e^{-i\phi/4}).\label{relation}
\end{equation}
\par
For the QD chain with $t_l=t_c$ and
$E_l=E_0$, the eigenenergies are given by
$e_m=E_0-2t_c\cos(\frac{m\pi}{N+1})$ and $[\bm\eta]$
matrix is expressed as
\begin{eqnarray}
&&[\bm\eta]=\sqrt{\frac{2}{N+1}}\cdot\notag\\&&\left[\begin{array}{cccc}
\sin\frac{N^2\pi}{N+1} & \sin\frac{N(N-1)\pi}{N+1}&\cdots & \sin\frac{N\pi}{N+1}\\
\sin\frac{N(N-1)\pi}{N+1} & \sin\frac{(N-1)^2\pi}{N+1}&\cdots &\sin\frac{(N-1)\pi}{N+1}\\
\vdots & &&\vdots\\
\sin\frac{N\pi}{N+1}&\sin\frac{(N-1)\pi}{N+1}& \cdots&\sin\frac{\pi}{N+1}\\
\end{array}\right].\ \notag
\end{eqnarray}
It can be readily found that in the absence of magnetic flux, for the QD chain with odd(even) QDs all their even(odd)-numbered molecular states decouple from the leads. On the other hand, when a local magnetic flux is introduced with $\phi=2\pi$, the decoupling result will be changed. Namely, for the QD chain with odd(even)QDs all their odd(even)-numbered molecular states decouple from the leads.

\begin{figure}
\centering \scalebox{0.45}{\includegraphics{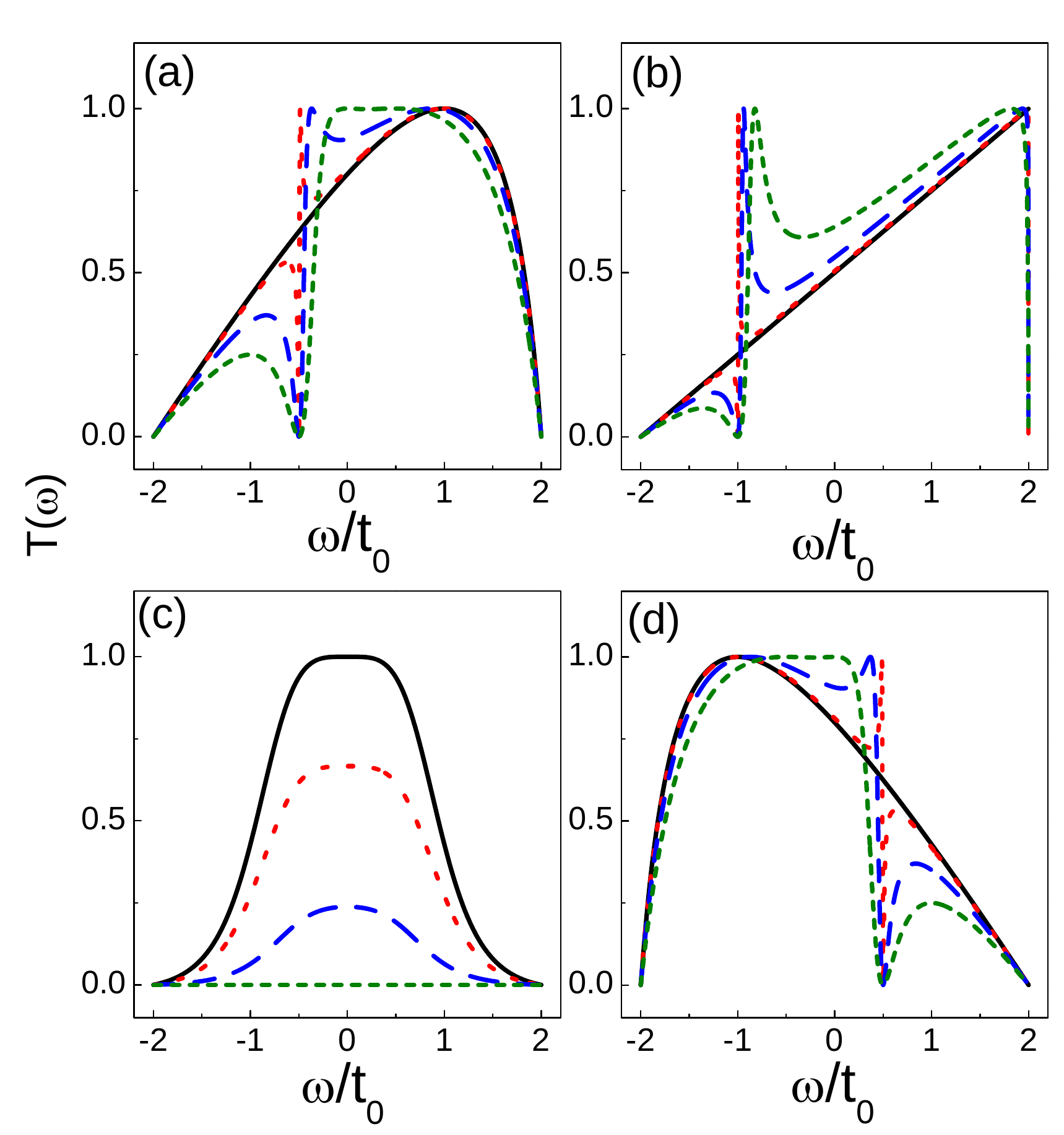}} \caption{
Spectra of the transmission function of the double-QD structure. (a)-(b) Results of $\gamma$=0, 0.1, 0.3, 0.5, in the cases of $t_c=0.5$ and $t_c=1.0$. (c)-(d) Results of $\phi$=$\pi$ and $2\pi$, in the case of $t_c=0.5$. \label{2QD}}
\end{figure}

\section{Numerical results and discussions \label{result2}}
Following the theory in the sections above, we proceed to perform the numerical calculation about the transmission function spectra of our considered system to clarify the influence of the $\mathcal{PT}$-symmetric complex potentials on the decoupling mechanism in the quantum transport process. Prior to the calculation, we take $t_0$ as the unit of energy.

\par
First of all, we investigate the transmission function spectrum in the double-QD geometry, i.e., $N=2$. Without loss of generality, we take the parameter values as
$t_{c}=0.5t_{0}$ and $v_0=t_0$ to perform the numerical
calculation. The corresponding results are shown in Fig.\ref{2QD}. It is clearly found that in the
absence of magnetic flux, the transmission function spectrum only shows one peak at the position of $\omega=t_c$. This result is exactly caused by the occurrence of the decoupling mechanism, and it can be analyzed as follows. In the case of $N=2$, the $[\bm\eta]$ matrix takes a form as $[\bm\eta]={1 \over\sqrt{2}}\left[\begin{array}{cc}
-1& 1\\
1& 1\\
\end{array}\right]$. With the help of Eq.(\ref{relation}) one can know that the coupling between the bonding
state $|1\rangle$ and lead-$\alpha$ $w_{\alpha
1}=v_0([\bm\eta]^\dag_{11}+[\bm\eta]^\dag_{21})$ is equal to zero, so the bonding state decouples from the leads completely. On the other hand,
when magnetic flux is introduced with $\phi=2\pi$, only one conductance peak appears corresponding to the
position of $\omega=-t_c$, as exhibited in Fig.\ref{2QD}(b). It is certain that the bonding state $|1\rangle$ couples to the leads and the antibonding state $|2\rangle$ becomes a bound state, because of $w_{\alpha m}=i v_0([\bm\eta]^\dag_{1m}-[\bm\eta]^\dag_{2m})$.
\par
Consider $E_1=E_0-i\gamma$ and $E_2=E_0+i\gamma$, we introduce the $\mathcal{PT}$-symmetric complex potentials to the two QDs of the QD chain to observe its influence on the quantum transport behavior. The results in Fig.\ref{2QD}(a)-(b) show that the nonzero $\gamma$ leads to the disappearance of the decoupling phenomenon. Meanwhile, distinct Fano lineshapes appear in the transmission function spectra, with the Fano antiresonance at the point of $\omega=-t_c$. It is additionally shown that the increase of $\gamma$ only widens the antiresonance valley but cannot change the antiresonance position. Next, if applying a local magnetic flux with $\phi=\pi$, we can find that $\mathcal{PT}$-symmetric complex potentials do not induce any antiresonance result but only suppresses the amplitude of the transmission function, as shown in Fig.\ref{2QD}(c). Only when the magnetic flux increases to $\phi=2\pi$, the non-Hermitian term causes the Fano antiresonance to appear at the position of $\omega=t_c$ [See Fig.\ref{2QD}(d)]. These results can be understood by writing out the analytical expression of the transmission function. To do so, we rewrite $T(\omega)$ in a form as $T(\omega)=|\tau_t|^2=|\sum_{jl}\tilde{v}_{Lj}G_{jl}\tilde{v}_{jR}|^2$ with $\tilde{v}_{\alpha j}=v_{\alpha j}\sqrt{\rho_0}$. After a straightforward deduction, we get the result that
\begin{small}
\begin{eqnarray}
\tau_t={2\Gamma_0\over \det\{{[G^r]}^{-1}\}}[(\omega-E_0)\cos{\phi\over2}+\gamma\sin{\phi\over2}+t_c]
\end{eqnarray}
\end{small}
with $\Gamma_0=v_0^2\rho_0$. And then,
\begin{small}
\begin{eqnarray}
&&\tau_t|_{\phi=2m\pi}=\notag\\
&&{2\Gamma_0[(\omega-E_0)\cos{m\pi}+t_c]\over [(\omega-E_0)^2-t_c^2]+\gamma^2-2\Sigma_{11}[(\omega-E_0)+t_c\cos{m\pi}]}.\notag
\end{eqnarray}
\end{small}
It clearly shows that the presence of $\cal PT$-symmetric complex potentials destroys decoupling mechanism and leads to the occurrence of antiresonance. To be concrete, in the case of $\phi=0$, the antiresonance occurs at the point of $\omega=E_0-t_c$, whereas it takes place at the position $\omega=E_0+t_c$ when $\phi=2\pi$, irrelevant to the increase of $\gamma$. We can additionally find that for a nonzero $\gamma$, $[\bm\eta]={1 \over\sqrt{2}}\left[\begin{array}{cc}
1& 1\\
-e^{-i\theta}& e^{i\theta}\\
\end{array}\right]$ with $\theta=\arctan{\delta\over t_c}$.
As a result, in the absence of magnetic flux, the magnitude of $v_{\alpha 1}$ is much smaller than that of $v_{\alpha 2}$. This provides the necessary condition for the occurrence of Fano effect.
\begin{figure}[htb]
\centering \scalebox{0.45}{\includegraphics{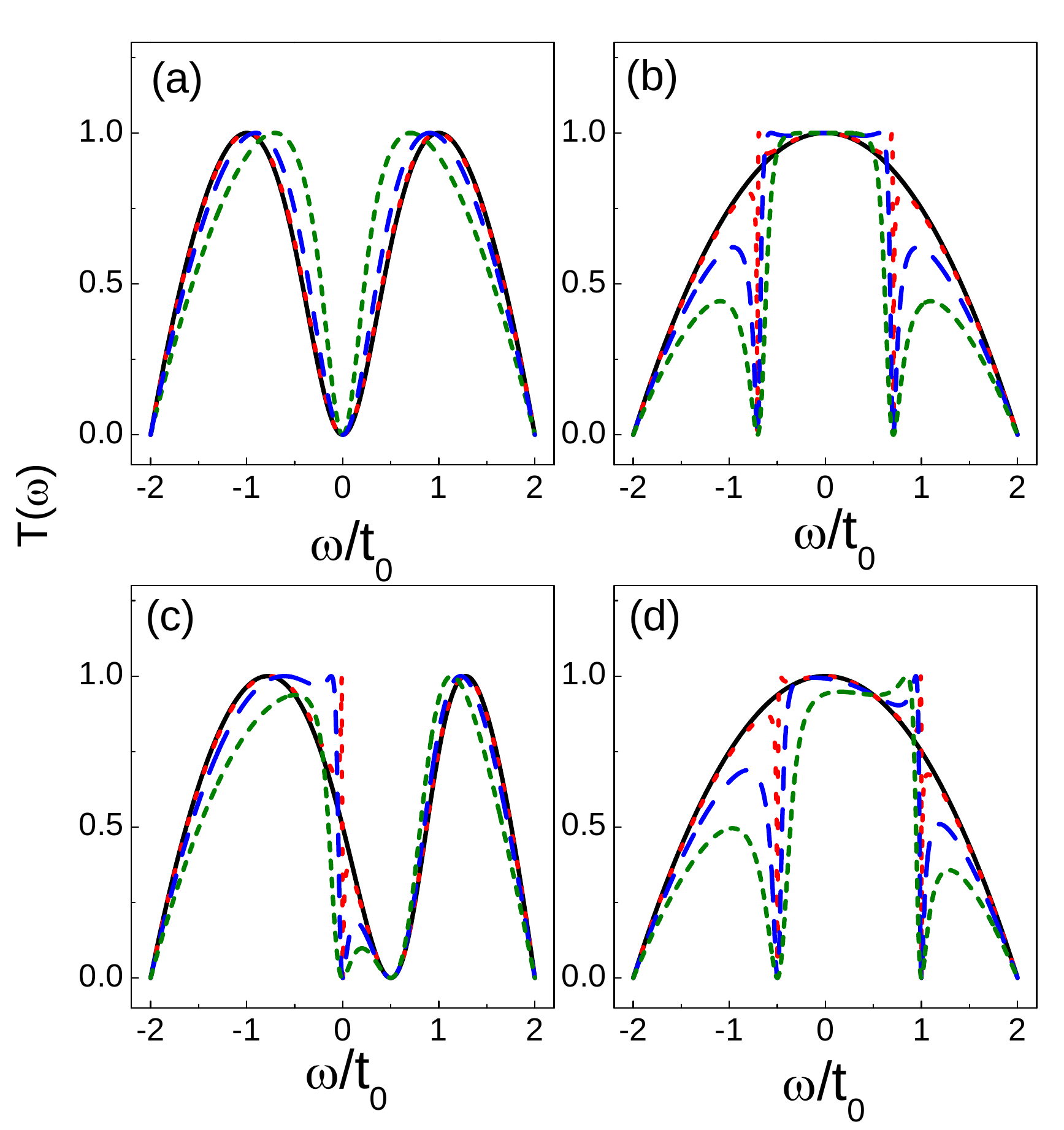}} \caption{
Spectra of the transmission function of the triple-QD chain. The structural parameters are taken to be $E_0=0$ and $t_c=0.5$. (a)-(b) Results of $\gamma$=0, 0.1, 0.3, 0.5, in the cases of $\phi=0$ and $2\pi$ with $E_2=0$. (c)-(d) Results of $\gamma$=0, 0.1, 0.3, 0.5, in the cases of $\phi=0$ and $2\pi$ with $E_2=0.5$. \label{3QD}}
\end{figure}
\par
Based on the double-QD results, we increase the QD number of the QD chain to $N=3$, to discuss the decoupling mechanism modified by the $\mathcal{PT}$-symmetric complex potentials. The numerical results are shown in Fig.\ref{3QD}.
We first find that in the case of $E_l=E_0$, only two peaks exist in the transmission function curve, corresponding to the positions of
$\omega=E_0\pm\sqrt{2}t_c$ respectively. At the point of $\omega=E_0$, no peak can be observed but the transmission function encounters its zero [see the solid line in
Fig.\ref{3QD}(a)]. By solving the $[\bm\eta]$ matrix
and using the relation $w_{\alpha
m}=v_0([\bm\eta]^\dag_{1m}+[\bm\eta]^\dag_{Nm})$, one can find in such
a case $v_{\alpha2}$ is always equal to zero, which causes the second molecule state $|2\rangle$ to decouple from the leads. When finite magnetic flux is taken into account with $\phi=2\pi$, the transmission function spectrum show up as one Breit-Wigner lineshape with its peak at the point of $\omega=0$, as shown in Fig.\ref{3QD}(b). Such a result should be attributed to the decoupling of the first and third molecular states from the leads.
In Fig.\ref{3QD}(a)-(b), we can also observe the interesting influence of the $\mathcal{PT}$-symmetric complex potentials on the quantum transport. First, as shown in Fig.\ref{3QD}(a), the nonzero $\gamma$ only narrows the antiresonance valley around the energy zero point but does not induce any other phenomenon. When the magnetic flux increases to $\phi=2\pi$, the non-Hermitian term eliminates the decoupling phenomenon and leads to the Fano antiresonance at the position of $\omega\approx\pm0.7$.
\par
Due to the speciality of the $N=3$ geometry, we can find that even in the case of $E_2=E_0+\delta$, the decoupling result still survives because
\begin{eqnarray}
&&[\bm\eta]=\frac{1}{\sqrt{2\Delta}}\left[\begin{array}{ccc}
\frac{2t_c}{\sqrt{\Delta-\delta}}&\sqrt{\Delta-\delta}
&\frac{2t_c}{\sqrt{\Delta-\delta}}\\
\sqrt{\Delta}&0&-\sqrt{\Delta}\\
\frac{2t_c}{\sqrt{\Delta+\delta}}&-\sqrt{\Delta+\delta}
&\frac{2t_c}{\sqrt{\Delta+\delta}}\\
\end{array}\right]\ \notag
\end{eqnarray}
with $\Delta=\sqrt{\delta^2+8t_c^2}$. Meanwhile, since QD-2 is located at the system center, its level change cannot destroy the $\cal PT$ symmetry. Thus, we would like to investigate the quantum transport results in the case of $E_2=E_0+\delta$. As a result, we see in Fig.\ref{3QD}(c)-(d) that in the case of $\phi=0$ with $\delta=0.5$, the two peaks in the transmission function spectrum become different from each other, and one asymmetric lineshape comes up. This is exactly related to the change of $\bm\eta$ matrix. Besides, the antiresonance point shifts to the position of $\omega=0.5$. Next when $\phi=2\pi$, one Breit-Wigner lineshape forms in the transmission function spectrum [See Fig.\ref{3QD}(d)]. If one nonzero $\gamma$ is taken into account, it causes a Fano antiresonance to appear at the energy zero point, with the antiresonance valley proportional to the value of $\gamma$. Alternatively, when magnetic flux increases to $\phi=2\pi$, two antiresonance points appear at the points of $\omega=1.0$ and $\omega=-0.5$, respectively. The antiresonance in these two cases exactly correspond to the molecular-state level, i.e., $e_1=E_0+{1\over2}(\delta-\sqrt{\Delta})$, $e_2=E_0$, and $e_3=E_0+{1\over2}(\delta+\sqrt{\Delta})$.

\par
Similar to the double-QD case, these results can be explained with the help of the analytical expression of $\tau_t$ and the matrix form of $\bm\eta$. For the triple-QD geometry with $E_{1(3)}=E_0$, we have
\begin{small}
\begin{eqnarray}
\tau_t=&&{2\Gamma_0\over \det\{{[G^r]}^{-1}\}}[(\omega-E_0)(\omega-E_2)\cos{\phi/2}\notag\\
&&+\gamma\sin{\phi/2}(\omega-E_2)+t_c^2(1-\cos{\phi/2})],
\end{eqnarray}
\end{small}
and
\begin{widetext}
\begin{eqnarray}
&&\tau_t|_{\phi=2m\pi}=\notag\\
&&{2\Gamma_0[(\omega-E_0)(\omega-E_2)\cos{m\pi}+2t^2_c\sin{m\pi\over2}]\over [(\omega-E_0)(\omega-E_2)-2t_c^2](\omega-E_0)+\gamma^2(\omega-E_2)-(\Sigma_{11}+\Sigma_{33})[\omega-E_0)(\omega-E_2)-2t_c^2\sin^2{m\pi\over2}]}.
\end{eqnarray}
\end{widetext}
One can clearly find that when $\phi=0$ the antiresonance occur at the points of $\omega=E_0$ and $\omega=E_2$, and they appear at the positions $\omega=E_0+{1\over2}(\delta\pm\sqrt{\Delta})$ when $\phi=2\pi$, which is also consistent with the molecular states of the QD chain and irrelevant to $\gamma$.
Now, we can have a preliminary understanding about the influence of the $\mathcal{PT}$-symmetric complex potentials. Firstly, they cause the ``previous" decoupled molecular states to re-contribute to the quantum transport via introducing the Fano antiresonances. Secondly, the $\mathcal{PT}$-symmetric complex potentials do not eliminate the ``previous" antiresonance mechanism.
\begin{figure}[htb]
\centering \scalebox{0.45}{\includegraphics{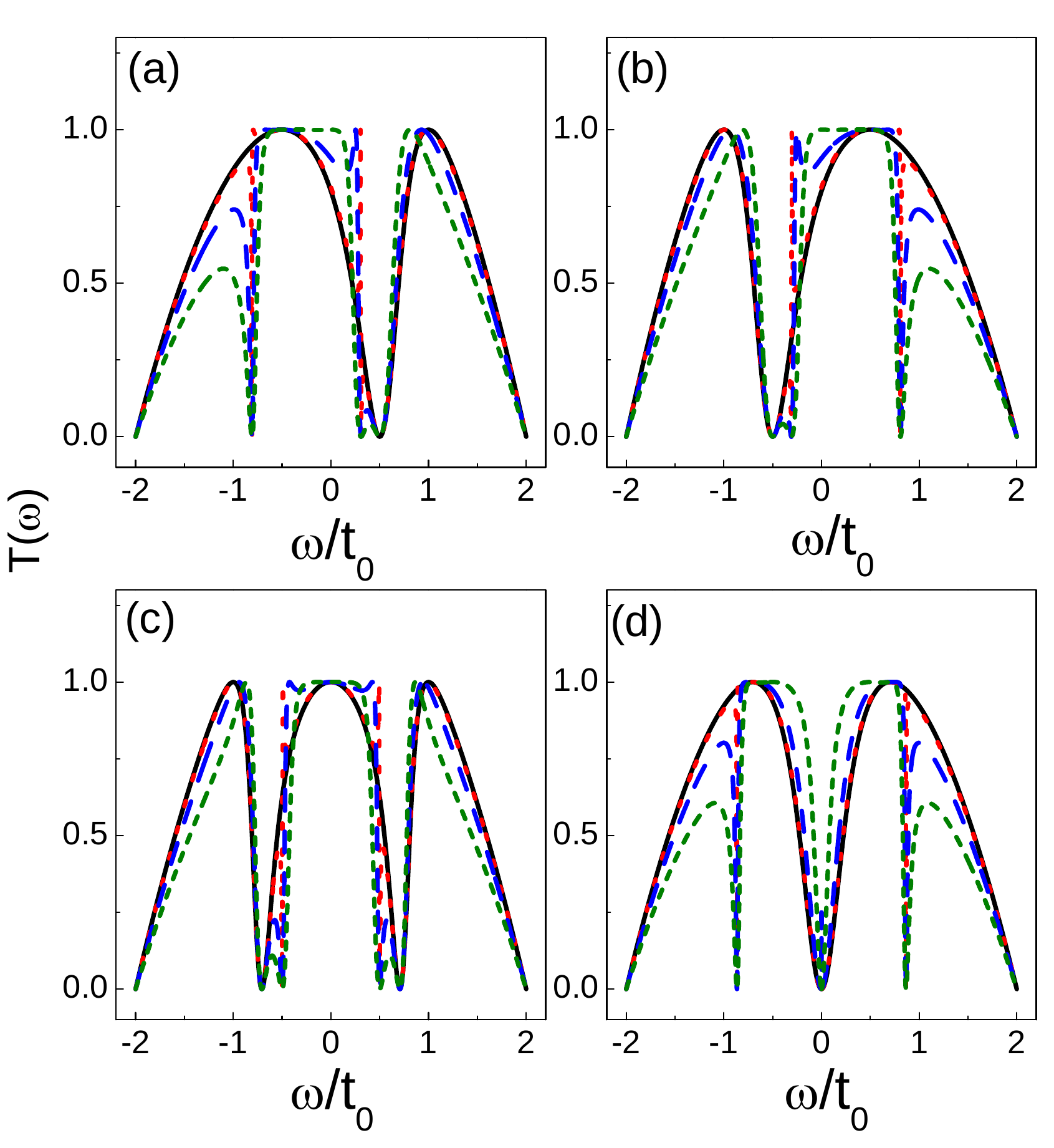}} \caption{
Transmission function spectra of the four-QD and five-QD chains. The structural parameters are taken to be $E_0=0$ and $t_c=0.5$. (a)-(b) Results of $N=4$ with $\gamma$=0, 0.1, 0.3, 0.5, in the cases of $\phi=0$ and $2\pi$. (c)-(d) $N=5$ results of $\gamma$=0, 0.1, 0.3, 0.5, in the cases of $\phi=0$ and $2\pi$. \label{45QD}}
\end{figure}
\begin{figure}[htb]
\centering \scalebox{0.40}{\includegraphics{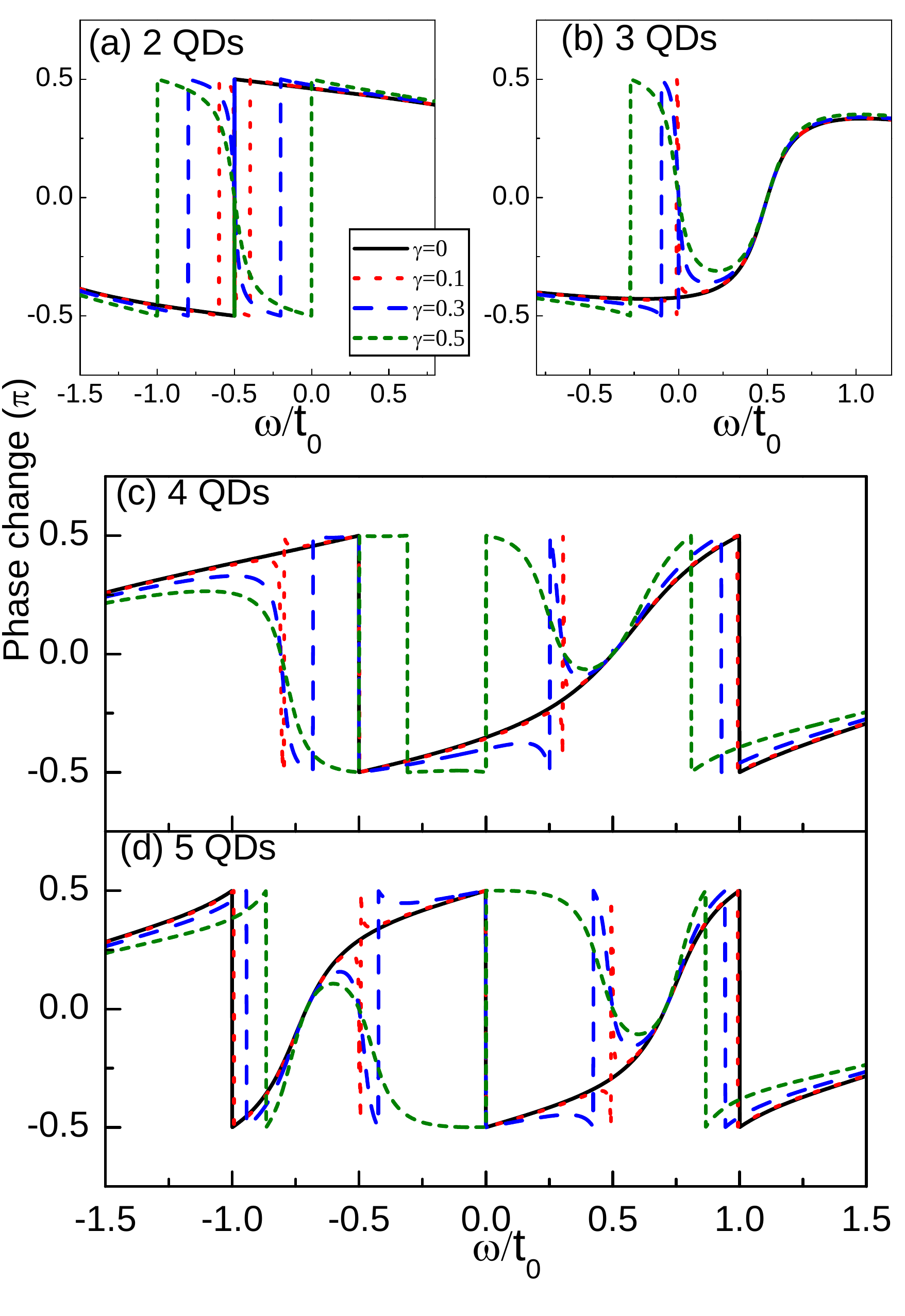}} \caption{ Phase of $\tau_t(\omega)$ in the case of zero magnetic flux. The QD number increases from 2 to 5, and $\gamma$ increases from zero to 0.5. \label{Phase}}
\end{figure}
\par
In order to check our conclusion above, we in Fig.\ref{45QD} plot the transmission function spectra of $N=4$ and $N=5$ geometries. In Fig.\ref{45QD}(a)-(b) where $N=4$, we see the remarkable decoupling phenomenon. Namely, in the absence of magnetic flux, the odd molecular states decouple from the leads. Meanwhile, apparent antiresonance occurs at the point of $\omega=t_c$ which exactly corresponds to the antibonding level of the sub-molecule of the chain without terminal QDs.
When the magnetic flux increases to $\phi=2\pi$, the even molecular states decouple from the leads, and antiresonance occurs at the point of $\omega=-t_c$ which is consistent with the bonding level of the sub-molecule of the chain without terminal QDs. Next, when the $\mathcal{PT}$-symmetric complex potentials are introduced to the terminal QDs, the decoupling phenomenon is found to transform into the Fano antiresonance, around the position of the ``previous" decoupled molecule states.
Next, Fig.\ref{45QD}(c)-(d) show the case of $N=5$ where $\phi=0$ and $\phi=2\pi$, respectively. It can be readily found that in the absence of magnetic flux, the even molecule states decouple from the leads. Meanwhile, apparent antiresonance occurs at the points of $\omega=\pm\sqrt{2}t_c$ which exactly corresponds to the odd-numbered levels of the sub-molecule of the QD chain. On the other hand, in the case of $\phi=2\pi$, the odd molecule states decouple from the leads, and antiresonance occurs at the point of $\omega=0$ which is consistent with the even-numbered level of the sub-molecule of the QD chain. As for the role of the $\mathcal{PT}$-symmetric complex potentials, it is to destroy the decoupling phenomenon and induce the occurrence of new antiresonance, identical with the results in the cases above. Based on these results, we can ascertain the quantum transport properties in such a structure.

\par
Quantum transport process is usually accompanied by the phase transition of the particle wave. Based on such an idea, in Fig.\ref{Phase} we plot the curve of the phase of $\tau_t(\omega)$ as a function of the incident-particle energy. Here we take the case of zero magnetic flux as an example to expand discussion. The structural parameters are the same as those in the above figures. In Fig.\ref{Phase}(a) where $N=2$, we see that in the case of $\gamma=0$, one sharp phase transition occurs at the point of $\omega=-0.5$. Next when the nonzero $\gamma$ is considered, two sharp phase transition processes can be found, and they are separated by the smooth phase transition at the point of $\omega=-0.5$. In comparison with the results in Fig.\ref{2QD}(a), we can know that these two sharp phase transitions originate from the appearance of the peak in the transmission function spectrum. Besides, one smooth phase transition takes place, corresponding to the antiresonance point in Fig.\ref{2QD}(a). Next, it shows in Fig.\ref{Phase}(b) that two smooth phase transitions appear the points of $\omega=0$ and $\omega=0.5$, respectively. They are exactly the antiresonance points for the triple-QD structure where $E_2=E_0+\delta$ with $\delta=0.5$. Similar phenomena can also be observed in Fig.\ref{Phase}(c)-(d). These results indicate that in our considered structure, the antiresonance is exactly consistent with the smooth phase transition.
\section{summary\label{summary}}
To sum up, we have presented a comprehensive analysis about the influence of $\mathcal{PT}$-symmetric complex potentials on the decoupling mechanism in a non-Hermitian system, by choosing one system where the terminal QDs of a QD chain couple equally to the left and right leads. After calculation, it has been found that in the absence of complex potentials, remarkable decoupling phenomenon has an opportunity to take place. To be specific, in the structure with odd(even) QDs, all their even(odd)-numbered molecular states decouple from the leads in the absence of magnetic flux. Meanwhile, apparent antiresonance occurs in transport through this system, which are related to the even(odd)-numbered eigenenergies of the sub-molecule of the chains without terminal QDs. In the presence of appropriate magnetic flux, such a phenomenon will be changed. Next, when the $\mathcal{PT}$-symmetric complex potentials are introduced to the terminal QDs, the decoupling phenomenon just transforms into the Fano antiresonance. We believe that the results in this work can assist to understand the interplay among the decoupling mechanism, antiresonance results, and $\mathcal{PT}$ symmetry in non-Hermitian discrete systems.
\section*{Acknowledgments}
This work was financially supported by the Fundamental Research
Funds for the Central Universities (Grant No.
N130505001), the Natural Science Foundation of Liaoning province of
China (Grant No. 2013020030), the Liaoning BaiQianWan Talents
Program (Grant No. 2012921078).

\clearpage

\bigskip


\begin{thebibliography}{99}

\bibitem{Bender1} C. M. Bender and S. Boettcher, Phys. Rev. Lett. \textbf{80}, 5243 (1998).

\bibitem{Bender3} C. M. Bender, D. C. Brody, and H. F. Jones, Phys. Rev. Lett. \textbf{89},
270401 (2002).
\bibitem{Mosta} A. Mostafazadeh, J. Math. Phys. \textbf{43}, 205 (2002); \emph{ibid} \textbf{43}, 2814
(2002).
\bibitem{Bender4} C. M. Bender, D. C. Brody, and H. F. Jones, Phys. Rev. D \textbf{70},
025001 (2004); H. F. Jones, J. Phys. A \textbf{39}, 10123 (2006).
\bibitem{Rotter} I. Rotter, J. Phys. A \textbf{42}, 153001 (2009).
\bibitem{Gold} I. Y. Goldsheid and B. A. Khoruzhenko, Phys. Rev. Lett. \textbf{80},
2897 (1998).
\bibitem{Hei} J. Heinrichs, Phys. Rev. B \textbf{63}, 165108 (2001).
\bibitem{Moli} L. G. Molinari, J. Phys. A \textbf{42}, 265204 (2009).


\bibitem{Klaiman} S. Klaiman, U. G\"{u}nther, and N. Moiseyev, Phys. Rev. Lett.
\textbf{101}, 080402 (2008); K. G. Makris, R. El-Ganainy, D. N.
Christodoulides, and Z. H. Musslimani, \emph{ibid.} \textbf{100}, 103904
(2008).
\bibitem{Xu} A. A. Sukhorukov, Z. Xu, and Y. S. Kivshar, Phys. Rev. A \textbf{82},
043818 (2010).
\bibitem{Kottos} H. Ramezani, D. N. Christodoulides, V. Kovanis, I. Vitebskiy,
and T. Kottos, Phys. Rev. Lett. \textbf{109}, 033902 (2012).
\bibitem{Makris} Z. H. Musslimani, K. G. Makris, R. El-Ganainy, and D. N.
Christodoulides, Phys. Rev. Lett. \textbf{100}, 030402 (2008).
\bibitem{Luo} X. B. Luo, J. H. Huang, H. H. Zhong, X. Z. Qin, Q. T. Xie, Y.
S. Kivshar, and C. H. Lee, Phys. Rev. Lett. \textbf{110}, 243902 (2013).


\bibitem{Hu} Y. C. Hu and T. L. Hughes, Phys. Rev. B \textbf{84}, 153101 (2011).
\bibitem{Zhu} B. G. Zhu, R. L\"{u}, and S. Chen, Phys. Rev. A \textbf{89}, 062102 (2014).

\bibitem{Guo} A. Guo, G. J. Salamo, D. Duchesne, R. Morandotti, M. Volatier-
Ravat, V. Aimez, G. A. Siviloglou, and D. N. Christodoulides,
Phys. Rev. Lett. \textbf{103}, 093902 (2009).
\bibitem{Kip} C. E. R\"{u}ter, K. G. Makris, R. El-Ganainy, D. N. Christodoulides,
M. Segev, and D. Kip, Nat. Phys. \textbf{6}, 192 (2010).


\bibitem{Miri} A. Regensburger, M. A. Miri, C. Bersch, J. N\"{a}ger, G.
Onishchukov, D. N. Christodoulides, and U. Peschel, Phys. Rev.
Lett. \textbf{110}, 223902 (2013).

\bibitem{Li} J. Schindler, A. Li, M. C. Zheng, F. M. Ellis, and T. Kottos,
Phys. Rev. A \textbf{84}, 040101(R) (2011).

\bibitem{Zo} M. Znojil, Phys. Rev. A \textbf{82}, 052113 (2010).

\bibitem{Bendix} O. Bendix, R. Fleischmann, T. Kottos, and B. Shapiro,
Phys. Rev. Lett. \textbf{103}, 030402 (2009).
\bibitem{Jin} L. Jin and Z. Song, Phys. Rev. A \textbf{80}, 052107 (2009); W. H. Hu,
L. Jin, Y. Li, and Z. Song, \emph{ibid.} \textbf{86}, 042110 (2012).
\bibitem{Jogle} Y. N. Joglekar and A. Saxena, Phys. Rev. A \textbf{83}, 050101(R)
(2011).
\bibitem{Valle} G. D. Valle and S. Longhi, Phys. Rev. A \textbf{87}, 022119 (2013).


\bibitem{Datta} S. Datta, \emph{Electron Transport in Mesoscopic Systems}, Cambridge
University Press, Cambridge, UK, 1997.
\bibitem{Fano} J. G\"{o}res, D. Goldhaber-Gordon, S. Heemeyer, M. A. Kastner,
H. Shtrikman, D. Mahalu, U. Meirav, Phys. Rev. B \textbf{62}, 2188 (2000); I. G. Zacharia, D. Goldhaber-Gordon, G. Granger, M. A. Kastner, Y. B. Khavin, H. Shtrikman, D. Mahalu, U. Meirav, Phys. Rev. B \textbf{64}, 155311
(2001).

\bibitem{Decouple} K. Bao and Y. Zheng, Phys. Rev. B \textbf{73}, 045306 (2005);
\bibitem{Kondo} N. J. Craig, J. M. Taylor, E. A. Lester, C. M. Marcus, M. P.
Hanson, and A. C. Gossard, Science \textbf{304}, 565 {2004}.
\bibitem{Fano2} K. Kobayashi, H. Aikawa, S. Katsumoto, Y. Iye, Phys. Rev. Lett. \textbf{88}, 256806
(2002); K. Kobayashi, H. Aikawa, S. Katsumoto, Y. Iye, Phys. Rev. B \textbf{68}, 235304
(2003); M. Sato, H. Aikawa, K. Kobayashi, S. Katsumoto, Y. Iye, Phys.
Rev. Lett. \textbf{95}, 066801 (2005); K. Kobayashi, H. Aikawa, A. Sano, S. Katsumoto, Y. Iye, Phys.
Rev. B \textbf{70}, 035319 (2003).
\bibitem{PRA1} B. Zhu, Rong L\"{u}, and S. Chen, Phys. Rev. A \textbf{91}, 042131 (2015).
\bibitem{Lvrong} Q. B. Zeng, S. Chen, R. L\"{u}, arXiv:1608.00065.
\bibitem{GF1} Y. Meir and N. S. Wingreen, Phys. Rev. Lett. \textbf{68}, 2512 (1992);
A. P. Jauho, N. S. Wingreen, and Y. Meir, Phys. Rev. B \textbf{50}, 5528 (1994).
\bibitem{GF2} W. J. Gong, X. Y. Sui, Y. Wang, G. D. Yu and X. H. Chen, Nanoscale Research Lett. \textbf{8}, 330 (2013); W. J. Gong, S. F. Zhang, Z. C. Li, G. Yi, and Y. S. Zheng, Phys. Rev. B \textbf{89}, 245413 (2014).
\bibitem{Self} S. Zhang, W. Gong, G. Wei, and A. Du, J. Appl. Phys. \textbf{109}, 023704 (2011).
\end{thebibliography}
\end{document}